# The Sky Above The Clouds

A Berkeley View on the Future of Cloud Computing

Sarah Chasins, Alvin Cheung, Natacha Crooks, Ali Ghodsi, Ken Goldberg, Joseph E. Gonzalez, Joseph M. Hellerstein, Michael I. Jordan, Anthony D. Joseph, Michael W. Mahoney, Aditya Parameswaran, David Patterson, Raluca Ada Popa, Koushik Sen, Scott Shenker, Dawn Song, Ion Stoica

**Executive Summary:** *Technology ecosystems often undergo significant transformations as they mature. For example, telephony, the Internet, and PCs all started with a single provider, but in the United States each is now served by a competitive market that uses comprehensive and universal technology standards to provide compatibility. This white paper presents our view on how the cloud ecosystem, barely over fifteen years old, could evolve as it matures.*

*Each of the early entrants to the cloud computing market offered their own proprietary interfaces. We predict that this market, with the appropriate intermediation, could evolve into one with a far greater emphasis on compatibility, allowing customers to easily shift workloads between clouds. However, the cloud's trajectory towards this more compatible and competitive market will be quite different from the examples cited above. In particular, we believe that a comprehensive compatibility standard supported by all clouds—essential in creating the markets in telephony, the Internet, and PCs—is neither likely to emerge nor necessary to allow customers to move workloads between clouds, and would in fact impede innovation. Instead, we argue that to achieve this goal of flexible workload placement, cloud computing will require intermediation, provided by systems we call* intercloud brokers*, so that individual customers do not have to make choices about which clouds to use for which workloads, but can instead rely on brokers to optimize their desired criteria (e.g., price, performance, and/or execution location). We believe that the competitive forces unleashed by the existence of effective intercloud brokers will create a thriving market of cloud services with many of those services being offered by more than one cloud, and this will be sufficient to significantly increase workload portability.*

*We think this compatibility-enhancing maturation of the cloud market will bring several important benefits, including: (i) lower barriers to cloud usage, thereby expanding the overall cloud market; (ii) rapid technical innovation via specialized clouds (which will give users access to best-of-breed services and hardware); (iii) more complete integration of various computational options (e.g., edge computing, on-premise computing, and choice of availability zones within individual clouds); and (iv) opportunities for enhancing compliance, security, and resilience via cross-cloud deployments (e.g., the hosting of model inference in multiple clouds to improve availability, or processing confidential data wherever needed to satisfy new regulatory constraints such as data and operational sovereignty). However, we do not see this particular path to compatibility as an inevitable outcome of maturation, but as a desirable and attainable possibility that requires the introduction of intercloud brokers to become a reality. Thus, this white paper is not just a passive description of a proposed technical innovation, but a call for its creation. Specifically, we are asking researchers and practitioners to join in building an early version of an intercloud broker that will help engender a new future for cloud computing, which we call "Sky Computing."*

# S1: Background

Technology ecosystems are often shaped by their origins. For instance, communication platforms like telephony and the Internet are built around comprehensive and universal standards that ensure interprovider compatibility, since such platforms were always intended to reach all users. In contrast, the cloud arose as a more cost-effective replacement for on-premise computing, and thus only a single customer at a time was involved. As a result, the early cloud providers did not seek cross-platform compatibility and instead offered proprietary interfaces. These proprietary interfaces, and other measures such as volume discounts and much higher data-transfer fees for egress than ingress, are now part of a broader strategy to lock customers into their current cloud provider.

Cloud computing is now a critical component of our increasingly interconnected IT infrastructure, which is causing almost all ecosystem actors—corporate users, third-party software services, and even many cloud providers—to seek more cross-cloud compatibility, each with their own motivations. For instance, corporate users have several distinct reasons for wanting the ability to shift workloads between clouds. Most urgently, they do not want such an essential component of their business being tied to one cloud provider, because they lose commercial leverage and have no protection when that provider suffers a major outage. They also need to comply with the growing number of government regulations that restrict where computation is carried out and personal data can reside, including strict regulations about data and operational sovereignty.[1] If a customer needs to expand to a new country where their cloud of choice doesn't have a datacenter, that customer will have no choice but to use another cloud that does. In addition, companies are increasingly interacting with other companies via the cloud to provide services or implement partnerships, which requires flexibility in workload placement. Thus, many companies see workload portability as moving from "nice-to-have" to "must-have" in the near future.

As cloud computing has grown, third-party companies—who provide a variety of cloud-based services such as data storage, data analytics, machine learning, video streaming, business intelligence, and identity management—are playing an increasingly important role in the cloud ecosystem. Greater cloud compatibility would enable these third-party software services to be more easily ported to multiple clouds, allowing them to reach additional users and giving them a competitive advantage over cloud-specific alternatives. Moreover, some corporate customers of these third-party services are already making cross-cloud functionality a requirement, for the reasons cited above.

Even many cloud providers have incentives for greater compatibility. In particular, clouds with smaller shares of the market—and as of this writing only two cloud providers have market shares significantly above 10% [Richter, 21]—want to increase compatibility so users can more easily shift workloads to them. Toward that end, they often offer service interfaces (such as for storage, container orchestration, and network management) that are compatible with (or at least similar to) the proprietary services offered by the more dominant clouds. Furthermore, a few clouds, including some large ones, support their interfaces on other clouds (e.g., Google's Anthos and Azure's Arc) to encourage developers to write applications against their APIs.

---

[1] These regulations require not only to process data within a country boundary but also in datacenters that are operated by the country nationals [OneTrust].

Along with these market forces creating a near-universal *desire* for greater compatibility, the widespread use of open source tools is providing additional *opportunities* for compatibility. As Section 2 describes, there are many open source tools available on most clouds, and they are now an important driver of cross-cloud compatibility.

Given the increasingly urgent demand for cross-cloud compatibility, and growing opportunities to achieve such compatibility through open-source tools, it seems clear that the cloud ecosystem must quickly move in this direction. However, what brought us here will not get us there. So, the question is: *what is the most effective way to help bring this change about?*

The usual tool for achieving compatibility in technology ecosystems is to universally adopt comprehensive standards, where the term comprehensive means that the standards are sufficient to ensure compatibility between providers. However, the calls for such standards in cloud computing have repeatedly failed (see Section 4), not due to lack of effort or insurmountable technical barriers but because there are few incentives for the dominant clouds to adopt such standards, as doing so would weaken their competitive advantage.[2] Furthermore, the API surface covered by the current clouds is enormous, with AWS alone offering over 200 services [AWS 21]. Standardizing such a huge number of APIs is a daunting if not impossible task. Indeed, history is littered with failed efforts to standardize large APIs, such as OSF/1 [OSF/1, 1993], which aimed to standardize the Unix operating systems in the 1990s, and CORBA, which aimed to standardize the way different applications and platforms share data [Henning, 2006].

More crucially, we think that adopting comprehensive standards for cloud services would impede innovation by locking the ecosystem into a set of interfaces that may not be appropriate for future uses. While standards in all technology ecosystems run this risk, we think this concern about premature and/or inappropriate standardization is particularly relevant to cloud computing. Users interact with the cloud through interfaces that range from low-level orchestration to high-level services (so it isn't clear which levels are appropriate for standardization) and change relatively rapidly over time (so it isn't clear when they are stable enough to standardize). We return to this issue in the next section.

This position paper advocates an alternate vision for creating a more compatible yet still evolvable cloud ecosystem.[3] The foundation of our vision is the use of "*intercloud brokers*."

---

[2] One exception to the failure of comprehensive standards is Globus which provides a unified identity and access management system, as well as a "fire-and-forget" secure and high performance tool to transfer and share files across different clouds. These services are complementary to Sky Computing, which can leverage them.
[3] The material here reflects a broader collaborative effort and a substantial evolution of our thinking from an earlier treatment in [Stoica and Shenker, 21].

Users of an intercloud broker specify their job characteristics and constraints, along with their desired optimization metrics (such as price or performance). Given this specification, an intercloud broker finds the best cloud on which to execute each portion of that job and then oversees its execution. The presence of intercloud brokers changes the fundamental cloud abstraction, with users interacting with a more coherent "Sky of Computing" rather than with a collection of individual and deliberately differentiated clouds. Furthermore, this proposal—which we call "*Sky Computing*"—imposes no requirements on clouds, but instead leverages existing market forces to encourage their active participation.[4]

In fact, even the dominant clouds might see Sky Computing as ultimately in their interest, as its increased ease of use—which transforms the cloud experience from a complex one (where users obtain resources from a cloud, configure those resources in a cloud-specific manner, and then execute a set of tasks on those resources) to a much simpler one of submitting job descriptions to an intercloud broker—might lead to a rapid expansion of the overall cloud market.[5]

In the next section, we more fully describe what Sky Computing is, but here we quickly clarify what it is not. Most importantly, Sky Computing is not a panacea. It does not magically make every workload seamlessly portable, but instead creates a market dynamic where an increasingly large class of workloads become sufficiently portable so that the concerns mentioned above (e.g., the need to obey various regulations, the ability to leverage best-of-breed services across clouds, and avoiding reliance on a single cloud provider) are addressed. This change will occur through co-evolution; cloud offerings will become increasingly compatible, and workloads will evolve so that they can leverage this compatibility.

In addition, Sky Computing is not merely an attempt to implement "*multi-cloud*" as it is currently envisioned, which typically involves running different workloads on different clouds. Instead, Sky Computing abstracts away clouds with intercloud brokers that can run the same workload on multiple clouds, or split individual jobs across different clouds, both of which are harder and more beneficial than traditional multi-cloud (see Inset A for a description of multi-cloud and a comparison with Sky Computing). While Sky Computing is more than multi-cloud, it is less than a call for comprehensive compatibility, by which we mean the ability to run all jobs on all clouds. Instead, Sky Computing will only achieve partial compatibility, where many jobs can run on multiple but not all clouds, with this compatibility growing over time. As we argue later, we think partial compatibility is preferable to full compatibility, as partial compatibility is sufficient for creating a competitive market while avoiding the rigidity of universal standards that would stifle innovation.[6]

---

[4] As we explain in Section 4, the terms Sky Computing and intercloud have been used before in different contexts, but have largely fallen into disuse, so we are reclaiming these apt words for our proposal.
[5] According to [Evans, 21], in 2021 only 15%-20% of corporate workloads were executed in the cloud, so a rapid expansion of the overall market would easily offset some decrease in market share.
[6] Here and in what follows, it is important to differentiate service-specific standards, such as a particular version of an open-source tool, from standards imposed on clouds for what services they must support, which is what we mean when we refer to comprehensive and universal standards.

Most importantly, even though Sky Computing is technically well within our reach, it is most definitely not inevitable. To achieve the promise of Sky Computing, we must (i) construct an open, effective, and *fine-grained two-sided market for cloud computing* through intercloud brokers, (ii) ensure that the market does not devolve in various dysfunctional ways, such as being thwarted by predatory pricing or collusion, and (iii) identify some "killer apps" that will drive its adoption. Thus, this document should not be read as a passive prediction of the future, a mere gazing into a cloudy crystal ball. Instead, we are describing a plan of action—for researchers and practitioners alike—for purposefully transforming the current cloud ecosystem into one that—by enabling freer competition and more rapid innovation through the use of intercloud brokers—better meets the needs of users, and of society more generally.

## S2: Definition

We now define Sky Computing by describing *what* it is, *how* it works, and *who* is involved. In the following sections we explain *why* we believe in this vision (Section 3) and then (Section 4) discuss various *objections* to our proposal, *risks* that the Sky might face, and *opportunities* that the Sky might create.

*What is Sky Computing?* Sky Computing is merely cloud computing that is mediated by an intercloud broker (see [Figure 1](#)). In Sky Computing, rather than users directly engaging with a particular cloud, users send a job and its description to an intercloud broker, which then selects the clouds on which various parts of the job are run and then manages their execution. Thus, an intercloud broker creates a two-sided market between users who offer jobs, and clouds that offer services. Many of these services (e.g., Kubernetes, Apache Spark, Apache Kafka) are offered by multiple clouds, while others are cloud specific (e.g., AWS Inferentia, BigQuery). We expect that there will be multiple intercloud brokers, some possibly specialized for different workloads.

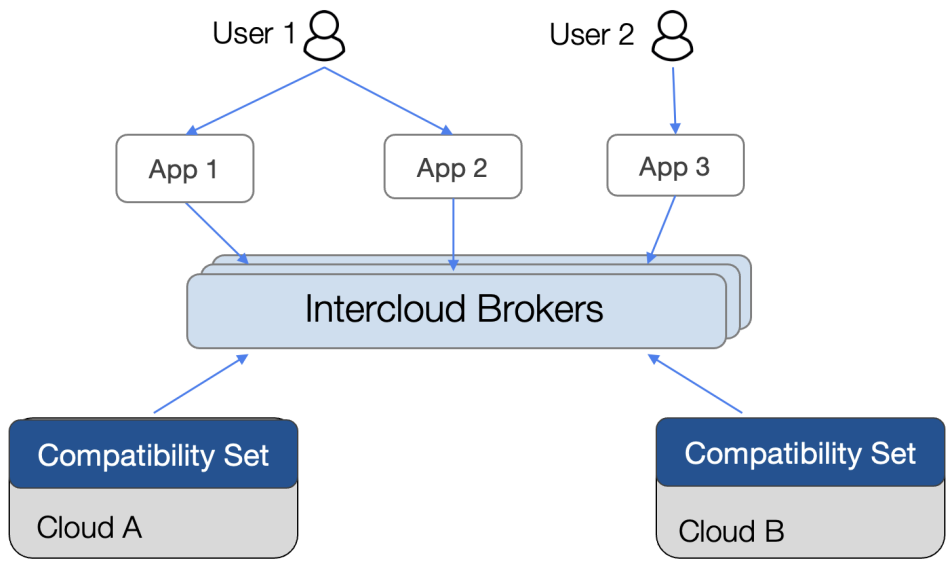

**Figure 1**: *Ths Sky Computing architecture.*

To make this description more concrete, consider the example (described in greater detail in Inset B) of a machine learning pipeline consisting of three stages: data processing, training, and model serving. Assume that the input data contains confidential information and that the user's goal is to minimize cost. Given this information, an intercloud broker could then run data processing on Azure, leveraging Azure Confidential Computing to anonymize the confidential data (or just remove the PII data), train on GCP to take advantage of TPU's cost-performance advantage for training, and serve on AWS to take advantage of Inferentia's lower costs for serving. Even with current egress charges and data transfer delays, splitting the computation across clouds can provide lower cost and faster processing by factors of two to three relative to running the pipeline on a single cloud. See Inset D for additional examples; we think that such examples will become more common as high-speed data exchange services such as Cloudflare's R2 lower both egress charges and data transfer delays.

*How does this work?* To perform this task, an intercloud broker needs the following set of functional components (see Figure 2):
- Service catalog: This catalog is a list of the service interfaces in the cloud ecosystem, along with the set of clouds that support each service. Each (service,cloud) entry contains instructions on how the service can be instantiated and managed on that cloud along with some price and performance information. The broker controls the information in the service catalog, but it can be based on input from clouds and various third parties.
- Job API: This API allows users to specify their job (where the input data resides, what processing is required, how data flows, etc.) along with their optimization metrics (e.g., cost or delay) and constraints (e.g., process data within the borders of a particular country, or within a given response time).
- Optimizer: Given the user's job specifications and requirements, the optimizer generates the optimal physical execution plan (e.g., selecting which clouds, instance types and number of instances,[7] and perhaps which locations within those clouds). The execution plan might also involve moving data to a different region in the same cloud or another cloud.
- Data orchestration: There are two ways to process data stored in a particular cloud region: run the computation stage in the same region or run the stage in a different region (and possibly cloud) but first move the data to that region. Data orchestration is in charge of moving or replicating the data to the region where the optimizer has decided to run the stage (note that the optimizer has factored in the cost and delay of such movement)
- Provisioner: The provisioner is responsible for allocating the resources across clouds to execute the physical plan. If the provisioner fails to allocate the resources on the cloud

---

[7] If the physical execution plan leverages a high level service like Snowflake for executing nodes in the DAG, it might not need to specify instance types and number of instances as these will be automatically determined by the service provider.

(or region) as specified by the physical plan (e.g., because the required resources are not available), it might ask the optimizer to generate another physical plan.
- Executor: The executor orchestrates the physical computational plan on the resources allocated by the provisioner, detecting and restarting failures and providing some degree of troubleshooting. A broker might use its own custom executor component or one of the commercially available orchestration and monitoring services.
- Billing service (optional): There is a range of possibilities for how customers are billed for using the Sky, and we expect several different models will coexist. In some cases, each cloud will bill the users directly for the portion of the job that they ran. In other cases, an intercloud broker could offer its own billing service, in which case it would have contracts with each cloud provider, pay for the job execution on behalf of the user, and then charge the user an aggregate fee.
- Identity and access management (IAM): This component will vary widely depending on the nature of the job, but at the very least the user must provide input data to the job, and that will likely involve passing some token or other form of credentials to the intercloud broker. Sky could also leverage some of the existing multi-party identity protocols and services, including Kerberos, Globus Auth, or commercial offerings such as Okta.

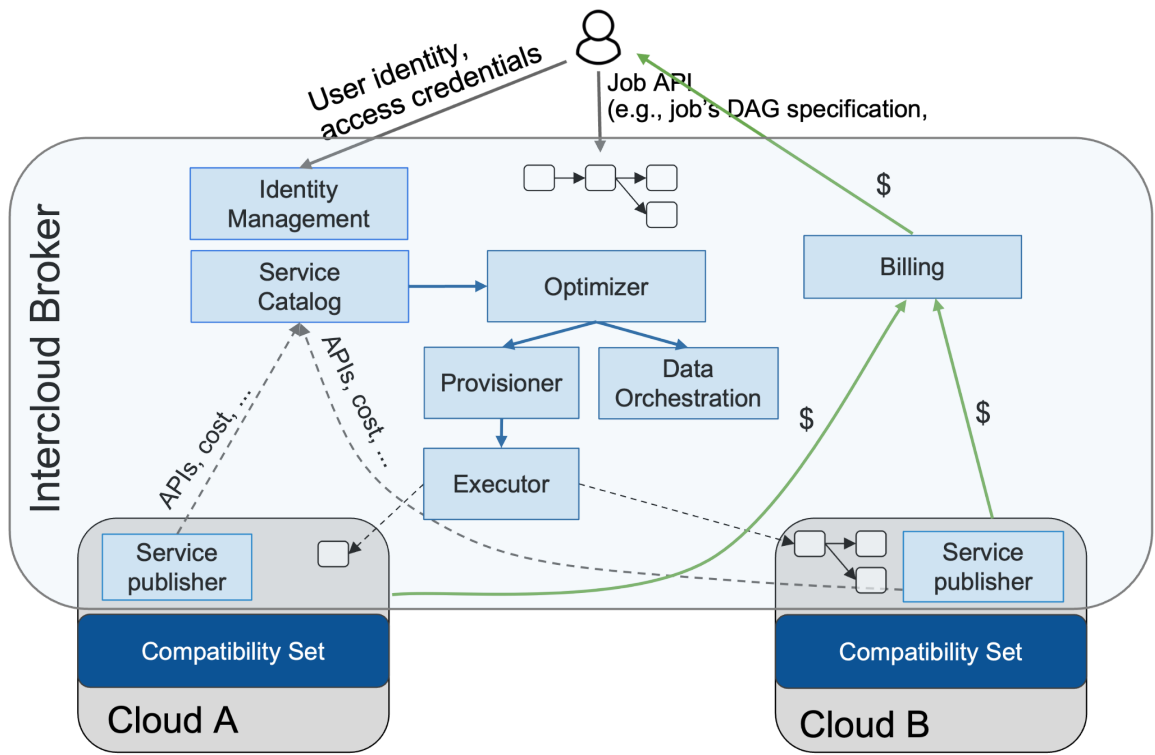

**Figure 2**: *The intercloud broker architecture*.

It will be difficult to build an initial intercloud broker whose components completely fulfill all these expectations. However, we see intercloud brokers evolving over time, starting with relatively simple versions that can only handle certain classes of workloads (such as batch jobs) and then expanding their scope over time. For instance, the job API should eventually handle a wide range of options for how processing is organized and how data flows, but to simplify the discussion that follows we assume that the initial version of the job API will describe a job as a directed acyclic graph (DAG), with specified services to be applied at each node in the DAG,[8] along with a description of what data flows to the next node in the DAG. In the simplest instantiation of this scenario, the IAM in this case needs only limit access to the submitting user, so sophisticated access controls are unnecessary. Furthermore, the initial version of the intercloud broker may not implement billing, and instead require users to have accounts with every cloud they want to use.

As Sky Computing advances, we will need significant advances in: how jobs and their requirements are specified in the job API; the way price, performance, and other relevant criteria are described in the service catalog; and how identity and access information is detailed in the IAM component. In addition, we should enable some degree of mix-and-match among these components (e.g., a particular broker might use a freely available service catalog but its own optimizer), which will require careful design of the interfaces between these components.

The service catalog contains essentially all services offered by the clouds, along with some price and performance information. In most cases the interfaces are defined by their underlying code, so the catalog would only list the relevant codebases and their versions (e.g., which versions of Apache Spark or Kubernetes). In the cases of widely adopted standards (e.g., SQL), the catalog would only list the relevant standards (e.g., SQL:2008), not each individual interface. We also expect there will be third party tests that an intercloud broker can use to verify the compatibility of these interfaces. For an open-source software based service, these tests could be the unit test suite of the API.

It is useful to define the "compatibility set" as those services supported by more than one cloud, and "proprietary services" as those not in the compatibility set. There can be competition for the portions of jobs that use interfaces in the compatibility set, while there is only a single cloud provider that supports proprietary interfaces. The service catalog includes both services in the compatibility set and proprietary services, and a "proprietary service" becomes part of the compatibility set when another cloud starts supporting that service. Thus, the degree of competition that Sky Computing engenders can be measured by size of the compatibility set, the breadth of support among clouds for the services therein, and the frequency of their use.[9]

---

[8] A node in the DAG can be an executable image (e.g., docker image, AMI), or can be be a source program or script (e.g., a Spark program) that can run on a service in the catalog (e.g., EMR, HDInsights, DataProc, Databricks).

[9] We want to emphasize that Sky Computing applies to the entire set of services, not just to the compatibility set, and in fact brings benefits even if all services are proprietary. In this latter case the benefit arises because Sky can run different stages of a job on different clouds, so even if some or all stages are proprietary, the intercloud broker can identify the appropriate clouds on which to run those stages. Our point here, though, is that the competition between clouds is enhanced if most job stages use services in the compatibility set.

Many of the services in the compatibility set are open source projects, either hosted by the clouds themselves or by third parties, and offer functionality such as cluster orchestration (e.g., Kubernetes), application packaging (e.g., docker), data processing (e.g., Flink), data ingestion (e.g., Kafka), and machine learning (e.g., Pytorch Lightning).[10] The interfaces in the compatibility set thus range from low-level (e.g., Kubernetes) to high-level (e.g., TensorFlow, SQL[11], Spark, Kafka) and everything in between (e.g., Redis, RocksDB). Lower-level interfaces provide developers with the greatest flexibility (e.g., you can run any code on Kubernetes), while high-level interfaces simplify application development while giving cloud providers more opportunities for more efficient implementations (e.g., highly performant implementations of SQL). We don't believe that there will be a single sweet spot in this tradeoff between user-flexibility and operator-innovation, so we expect that the compatibility set will always include many different levels of abstraction, and that even at the same level of abstraction interfaces will wax and wane in popularity as technologies and applications evolve over time (e.g., the rise and fall of Hadoop [Harrison, 2019]).

As an aside, we note that one approach to extending the compatibility set is to use shim layers.[12] A *shim layer* aims to provide a common API that abstracts away the differences between similar services already available on different clouds. Unfortunately, traditional shim layers have a significant drawback: they provide the lowest common denominator functionality across services. For example, a shim layer for existing cloud storage services offered by AWS, Azure, and GCP would not provide replication across different regions, despite the fact that both Azure's Blob Storage and GCP's Cloud Storage do, since AWS's S3 does not [Debjeet, 2021]. However, there is a concept closely related to shim layers we call bolt-on [Bailis, 2013], that can extend the functionality of existing services. Examples of bolt-on layers are Delta Lake, Iceberg, or Hudi that run on top of existing cloud storage services and provide support for fine grained updates and transaction semantics, which the underlying services do not. Thus, we expect bolt-on layers to be one of the ways proprietary services can be leveraged to create more widely supported services and thus grow the compatibility set.

Returning to our main argument, we think the diversity and dynamism of the compatibility set—properties quite different from a typical approach based on comprehensive and universal standards—are essential for the cloud ecosystem because of three features that together distinguish it from ecosystems where such comprehensive standards are appropriate.

First, we believe that any fixed and comprehensive set of cloud standards would, over time, significantly impede innovation. While most standards are suboptimal, many of the most successful ones are at such a low level that they leave plenty of room for innovation, and the suboptimality is often an acceptable tradeoff for achieving the widespread adoption (e.g., the IP protocol has many known flaws, but none of them prevented the flourishing of the Internet ecosystem and IP's universal standardization is essential to the Internet; similarly, the IBM PC design was far from optimal, but it was sufficient to create a thriving marketplace of PC applications, and PC-compatibility was necessary for it gaining market dominance). In these cases, one could identify a low-level compatibility layer that was sufficient to ensure compatibility while allowing innovation. We do not believe such a comprehensive compatibility layer exists for the clouds because, as discussed above, users want the ability to interact with clouds at many levels. It is not enough to just standardize on, say, Kubernetes, as a compatibility layer because many users want to invoke a higher-level interface (e.g., Apache Spark or Kafka); to ensure compatibility these higher-level interfaces must also be standardized. Freezing these higher-level interfaces in a standard would do great harm as they are continually modified to better serve customer needs.

---

[10] This can be available as docker images or as machine images provided by clouds themselves [here, here, here].

[11] One example is the SQL92 standard that is supported by many cloud-hosted database services, including Azure SQL, Google's Big Query, and AWS' RDS.

[12] The literature related to the concepts discussed in this paragraph has not arrived on an agreed-upon terminology, so here we use "shim" to mean the interposition of code that provides the lowest common denominator AP, and "bolt-on" to mean the interposition of code that provides an API with a higher level of functionality than the underlying systems.

Second, while for many standards-based ecosystems there is no such thing as partial compatibility (except in the most minor of features or extensions like IP options), the benefits of Sky Computing only require that many services can run on more than one cloud, so that intercloud brokers can establish a reasonably competitive market. Since there is no need for every service to run on all clouds, universal standards (by which we mean standards that all clouds must support) are an inappropriate solution to the cloud's compatibility problem.

Lastly, while we are far from having enough compatibility to make all workloads portable, we think there is currently enough compatibility for intercloud brokers to have an impact on the market, and that the market forces enabled by intercloud brokers can drive the ecosystem towards even greater levels of compatibility. To offset the lower levels of compatibility in the early stages of Sky Computing, the intercloud broker may often have to choose to move data to the clouds that support the desired services (rather than moving computation to the data). In general, it is easier to make data accessible on different clouds than porting or reimplementing a service on a new cloud. Existing cloud blob storages have similar interfaces (e.g, AWS' S3, Azure's Blob Storage, GCP's Cloud Storage), and popular open-source formats (e.g., Parquet) and data storage systems (e.g., [Delta Lake](), [Hudi](), [Iceberg]()) are already available on multiple clouds. Furthermore, there are already open-source protocols and services that allow applications to share the data across the clouds (e.g., Delta Sharing, Snowflake). Of course, moving data can incur non-trivial cost and latency overheads. The Optimizer will take into account these overheads when generating the job physical plan, and the Data Orchestration will aim to reduce these overheads.

Given these three considerations, we believe that for the cloud ecosystem a comprehensive and universal set of standards would (i) be infeasible (given the number of interfaces), (ii) be unadoptable (the dominant clouds would be resistant, as such standards would lessen their competitive advantage), and (iii) do more harm (stifle innovation) than good (because universal compatibility is not necessary for intercloud brokers to be effective). Thus, we should turn to the market rather than comprehensive and universal standards to create the desired level of compatibility.

*Who are the participants in Sky Computing?* Users can choose to use the intercloud broker interface (in which case they are using Sky Computing), or to interact directly with specific clouds (in which case they are not). In addition, users who want to derive maximum advantage from Sky Computing will use interfaces within the compatibility set, though brokers support all the services in the service catalog.

Individual clouds do not decide whether or not they participate in Sky Computing. Instead, they choose whether their business model focuses on differentiating their interfaces or on using interfaces in the compatibility set, with the latter choice allowing them to better compete for jobs mediated by intercloud brokers through better implementations or pricing of that service.[13] If they are more focused on interfaces in the compatibility set, they might also choose to offer reciprocally-free data peering, where a cloud A does not charge egress fees for data flowing to cloud B if in turn B does not charge egress fees for data flowing to cloud A. Such reciprocally-free peering lowers the cost of including such a cloud in the processing of a job, and thus improves its competitive positioning in the optimizer's analysis (at the cost of lost revenue of those egress fees). A cloud that focuses on compatible interfaces and engages in reciprocally-free data peering is embracing the Sky Computing vision more actively than one that does not, but intercloud brokers will send jobs to clouds regardless of their choices, as long as the cloud supports the required interfaces and is the best choice (in terms of the user's desiderata).

The last class of participants in Sky Computing are the entities that offer intercloud brokers. We are currently building an initial open-source prototype of an intercloud broker, as a proof-of-concept, but we assume that multiple providers who can make money by charging relatively small transaction fees will eventually provide such brokers. Thus, there will be competition among intercloud brokers, which will spur ongoing innovation in their implementations (e.g., leading to better pricing and performance information, and better monitoring and troubleshooting tools). Because there will be no formal standards for brokers in the beginning, these intercloud brokers can use reasonably effective but imperfect solutions to problems such as constructing a service catalog and collecting pricing information. By starting with a few targeted applications, rather than striving for full generality, they will further simplify the intercloud broker design and implementation. Some standards for brokers may eventually arise (e.g., for job APIs), but we note that standardizing broker interfaces is far different from standardizing what clouds must support. The latter directly controls what clouds must offer, while the former merely defines how we describe jobs, cloud offerings, user preferences, and the like.

## S3 Conjectures

---

[13] As we discuss later, we expect that in cloud computing as in all technology ecosystems, the current market leaders are likely to be lagging adopters of innovations that upset the current competitive landscape.

Sky Computing is a major shift in current cloud practices, and such shifts are rare in large ecosystems, so why do we feel optimistic about such a change happening? Our optimism rests on five conjectures. The first three are about how we might start the process of creating Sky Computing, the fourth is about how the market created by Sky Computing can lead to greater compatibility, and the fifth conjecture is about where that process of evolution might lead.

**Conjecture 1:** *Intercloud brokers need not initially handle all workloads; instead, starting with some easy but useful cases—including two conjectured "killer apps"—will be enough for Sky Computing to provide significant benefits to users.*

For Sky Computing to bring immediate benefits to users, a reasonably large class of jobs must meet three requirements. First, there must be sufficient compatibility so that these jobs have options as to where they are run. Second, there must be sufficient diversity in price and performance among these clouds so that users would care where their jobs run. Third, these jobs must fit the initial form of the job API, which we are presuming will be a DAG of processing nodes.

We believe that the cloud status quo meets all three requirements. Batch jobs typically fit the DAG model. Much of the compatibility for such jobs comes from the use of open-source projects. Inset C lists some of the open-source projects that are available across clouds. Inset D describes some examples where, due to variations in price and performance, Sky Computing would help meet user needs, thus supporting the conjecture that there are enough "easy but useful cases" to provide users with immediate and significant benefits from Sky Computing.

However, adoption of a new technology often depends on a "killer app" that provides the forceful impetus necessary for disturbing the status quo. Here we predict that there will be two "killer apps", one driven by a new opportunity and one driven by a new requirement. Data processing and machine-learning pipelines are a significant and rapidly growing component of today's cloud workloads. Since they fit the DAG model, and typically rely on open-source tools, they can be easily supported by the Sky, which (as we discussed above) has the potential for reducing both cost and delays for such jobs. We believe that these jobs will drive the initial wave of Sky usage by requiring little or no change to workloads yet producing significant cost and/or performance improvements.

In addition, companies handling private data will soon need to comply with data and operational sovereignty requirements. We believe that these regulations will come into effect faster than the clouds can install new datacenters in the countries imposing such regulations. As a result, companies will be forced to restructure their workloads to use services in the compatibility set, so that they can be shifted to clouds that have datacenters in the relevant countries. We believe that this use case will be the initial driver that causes companies to alter their workloads to adapt to Sky Computing.

These two trends, the increasing use of data processing and machine learning, and the increasing number of data and operational sovereignty regulations, show no sign of abating, and thus we predict that they will provide the two strongest initial drivers for Sky Computing.

**Conjecture 2:** *For clouds, it is enough to reward but not impose compatibility.*

We have repeatedly said that market forces will lead to Sky Computing, which might sound strange given that there is already strong competition between clouds. However, the market created by Sky Computing has two key differences from the current cloud market. First, it is at a much finer granularity, with intercloud brokers seeking out the best cloud for each stage of a job. Second, it is a two-sided market, with jobs being offered by users, and clouds implicitly (via the service catalogs in each intercloud broker) offering bids. The combination of these two factors—fine granularity and two-sidedness—transforms the competition between clouds from one where lock-in mechanisms, volume discounts, large sales forces, and extensive marketing campaigns play an essential role, to one where almost nothing matters but the ability of the cloud to best meet each user's needs. In short, Sky Computing allows users to leverage the "best-of-breed" for each stage in all their workloads.

Given such a market, we believe that many clouds, particularly those with smaller market shares, will find it advantageous to tailor their offerings to maximize revenue from Sky Computing. This means offering more services in the compatibility set, and perhaps specializing in certain classes of jobs where they have a competitive advantage (due to technology and/or pricing). Thus, our proposal of Sky Computing is not asking clouds to do anything other than act in their own immediate self-interest; initially, only the smaller clouds may find Sky Computing advantageous, but as more users use brokers to handle their interactions with the cloud, even some of the larger clouds may embrace Sky Computing more fully.

**Conjecture 3:** *An open-source version of intercloud broker is enough to get started.*

Sky Computing has a chicken-and-egg problem: who would build an intercloud broker before Sky Computing takes off, and how can Sky Computing take off without an effective intercloud broker? Solving this dilemma is where the research community and other interested parties can play an essential role, by creating a publicly available version of an intercloud broker that, while missing many features that will be needed in the long run, can provide immediate value now, at least for our two "killer apps" .

We are currently building such a system (see [Inset E](#) for a brief description) that is already capable of supporting some of the machine learning (ML) use cases mentioned in [Inset D](#). We will first make this system available to the research community (supporting use cases tailored to common research requirements) and then plan to offer it more broadly as we gradually expand its capabilities. One purpose of the document is to invite others to join in its design, development, and deployment; see the discussion in Section 4 about opportunities for research to have an impact on Sky's evolution. While the research community may not play an important role in the long-term economics of Sky Computing, it might be the only actor in the cloud ecosystem willing to take this critical first step to overcome the chicken-and-egg problem, and we will need your help to make this initial broker, or another initial prototype, a success. This is similar to what led to the Internet, which traces its roots back to work done by the research community in the 1970s [Cerf 93].

**Conjecture 4:** *Once initial intercloud brokers are deployed, time and the market will give rise to a more mature and capable version of Sky Computing.*

Intercloud brokers will naturally evolve through competition between broker providers to create an increasingly effective two-sided Sky Computing market. The presence of this two-sided market will create a self-reinforcing cycle of more compatibility (between clouds) and more workloads using the Sky (from users): See Figure 3. To be more specific, as clouds become more compatible, more of a user's workload can gain advantage from using the Sky. Similarly, as more jobs use the Sky, clouds are incentivized to offer more services in the compatibility set, so that they can compete for the submitted jobs. We think that once it begins, this process will gather momentum, creating an ever-increasing degree of compatibility while still allowing clouds to innovate by offering new services. Given this beachhead, the only issue is whether the critical ingredients for the Sky's initiation are present. They are (i) an initial version of an intercloud broker (which, as discussed in our third conjecture, we are building and invite others to help develop), and (ii) an economically meaningful set of jobs such as ML and data sovereignty that conform to the initial job API and that use at least some services in the compatibility set (which the first two conjectures address in the affirmative). Thus, we think time and the market will be sufficient to bring about a more mature and capable version of Sky Computing. The remaining question, then, is what kind of ecosystem will this eventually lead to, to which we now turn.

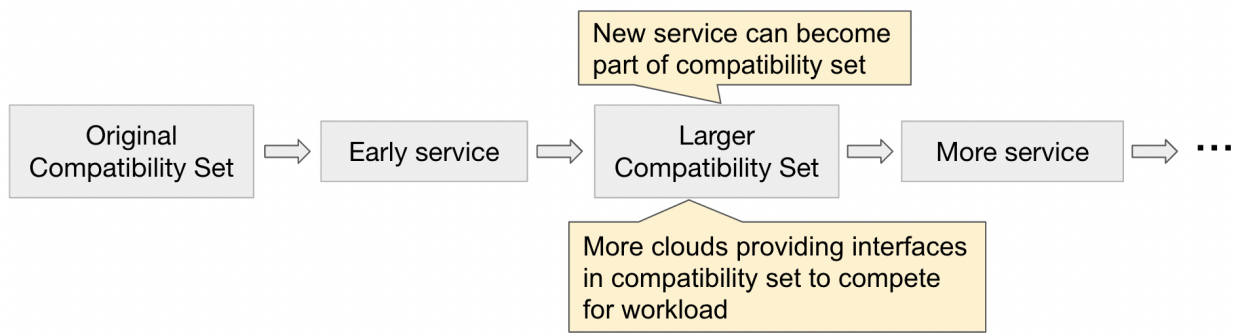

*Figure 3*: *Market forces creating a more mature and capable version of Sky Computing.*

**Conjecture 5:** *The Sky will lead to more specialized clouds, more rapid technical innovation, more complete integration of various computational options (e.g., edge computing and on-premise computing), an ongoing co-evolution of workloads and cloud practices, and more widespread cloud adoption.*

Clouds retaining a proprietary-service-oriented business model will attract those customers who think that large portions of their workloads are better off in one cloud using their proprietary interfaces. Clouds adopting a more Sky-oriented approach will attract customers for whom they can provide the best "offer" for at least some stages of their workloads. While there may be one or two large clouds that remain in the former category, for most other cloud providers trying to attract customers with their own proprietary offerings is unlikely to succeed. Thus, we expect that most cloud providers will orient their new offerings to maximize revenues from jobs mediated by the intercloud brokers.

What is notably different about this Sky-oriented business model is that some clouds can become quite specialized by, for example, providing hardware support for machine learning computations, or low-cost storage, or serverless offerings, while not offering the vast array of other services most current clouds offer. The recent announcement by Nvidia and Equinix [McPhee, 2021], and Cirrascale [Trader, 2021] suggest that this is already happening. Sky Computing allows specialized clouds to more easily find customers (i.e., they don't need marketing to reach customers, an intercloud broker will bring them), thereby making it easier for such specialized clouds to emerge. This specialization will speed innovation because (i) such infrastructures can be more finely tuned to particular use cases, (ii) technology companies can bring their products directly to market, rather than relying on one of the major cloud providers to deploy them, and (iii) prototypes developed by the research community can directly offer their services through brokers[14].

Note that with the Sky, the nature of the cloud ecosystem changes from a purely winner-take-all competition to a combination of direct competition (on the services they specialize in) and implicit cooperation (where each cloud relies on other clouds to support the services they don't). While this is quite different from the explicit standards-based federation seen in the Internet or cellular markets, it would represent a major shift in the cloud market, resulting in more shared interests and less direct competition.

The job API allows users to express various criteria to guide the placement of each node on the DAG. This naturally allows users to express various regulatory requirements (via requirements on location), to direct jobs to edge computing nodes (if low latency to clients is essential), and to direct jobs to their on-premise facilities (if low cost is paramount). This last option bears on the recent debate about the economics of cloud computing initiated by Wang and Casado's article [Wang and Casado, 2021] about the potentially massive cost savings of repatriating cloud workloads. We take no position on the substance of the debate, but note that Sky Computing could greatly ease such repatriation. This flexibility requires that repatriating workloads not be seen as "bringing workloads back on-premise" in a traditional sense. Instead, repatriating workloads should be seen as running a "private cloud," one which supports a set of interfaces that covers much of the institution's workload. Then repatriating merely involves setting a much lower cost for the private cloud, and letting intercloud brokers use the private cloud when appropriate.

We have focused on compatibility of computational interfaces, but there are many operational aspects of clouds—such as security policies, identity management, and network

---

[14] We assume that these research prototypes would initially be used mostly by the research community themselves, but might eventually gain the attention of commercial providers and/or users.

connectivity—that require detailed user configurations that are idiosyncratic to each cloud. The early versions of the job API will sidestep these complications by focusing on workloads where these configurations are straightforward (e.g., training on public data sets), but eventually Sky Computing must handle workloads where these operational requirements are less trivially accommodated. We think that such accommodations will arise from a co-evolution between workloads, intercloud brokers, and cloud providers where (i) brokers start supporting various examples or templates for these operational issues that capture common use cases (with the broker handling the translation from their template to each cloud's particular configuration settings), (ii) users start modifying their jobs, where possible, to fit these templates, and (iii) clouds modify their configuration systems to be compatible with such templates.

We have seen this evolution with *serverless computing*, which had very different operational requirements than what users had previously used and what clouds had previously supported, but both workloads and internal cloud operations changed to adapt to this innovation. We see a similar process happening with Sky Computing, where the initial offering does not address all of the user's current expectations nor fit the current operational practices of clouds. With Sky Computing, we see the co-evolution being led by the brokers as they define various templates, with users and clouds following by adopting and supporting such templates. Hence, we are not ignoring these operational issues; to the contrary, we see mastering them as crucial to the success of Sky Computing, but rather than seek immediate solutions to all of today's complexities (which we think is infeasible) we recommend an ongoing evolutionary approach of identifying common design patterns whose needs are not being met, and extending the broker's interfaces to handle such cases.

Lastly, we expect that the overall penetration of cloud computing will be accelerated by Sky Computing, because it makes moving workloads to the cloud much simpler than it is today. Our hope is that the current model of cloud usage, requiring extensive cloud-specific configuration (as discussed above) and explicit management of cloud resources, will largely give way to Sky's use of higher-level interfaces where most of these complications are hidden. This simplification may be the most important impact of Sky Computing, driving the level of abstraction up to the highest practical level and freeing users from most cloud-specific configurations.

## S4 Objections, Risks, and Opportunities

Every radical proposal is met with strenuous objections, and every technological development creates both unexpected opportunities and worrisome risks. Sky Computing is no different, and in this section we discuss Sky Computing's objections, risks, and opportunities.

### S4.1: Objections

We now review some objections to Sky Computing that we have heard, along with our brief responses.

**Objection**: *We have tried this before, and it failed.*
**Response:** There have been several attempts to create more compatibility among clouds. In fact, we are not the first to use the name "Sky Computing" in such a proposal; see the following papers, dating back to 2009, for early uses of this term [Keahey et al., 2009][Fortes et al., 2010][Monteiro et al., 2011]. However, these papers argue for a uniform API to be implemented on top of all clouds. One example is Nimbus, a cross-cloud Infrastructure-as-a-Service platform targeting specific workloads such as high-performance computing (HPC). In contrast, our proposal does not seek to impose a uniform API. Instead, we are enabling a two-sided market; clouds can decide to adopt interfaces in the compatibility set or not (and these will change over time), based on what is best for their business. The central thesis behind Sky Computing is that once this two-sided market is established, greater compatibility will emerge without it having to be imposed, and that the lack of fully comprehensive and universal standards will encourage innovation.

**Objection**: *Sky Computing will never be fully general.*
**Response:** We agree! One should think of Sky Computing not as a static endpoint, but as a process of meeting market demands. Sky Computing will start with the easy but useful cases, and then gradually expand the generality of its job API and other broker components. At every point in time, there will be jobs that Sky cannot handle, but if this class of jobs is economically important, the brokers will add interfaces to incorporate these requirements and clouds will add services (if needed) to support them. There are two points here. One, in many cases Sky will incorporate new classes of jobs not by producing an ever-more-general single interface, but by adding new job API interfaces for these new use cases; one can think of these as specialized broker interfaces, and not all brokers need to support all such interfaces. Second, it is unlikely that Sky will ever handle all use cases, but will continually absorb the most economically important use cases. Thus, the goal of Sky Computing is not to be fully general and handle every job, but to handle enough of the cloud workload so as to change the basic cloud ecosystem. Thus, when evaluating Sky Computing, it is more relevant to focus on what it can do, not what it can't.

**Objection**: *Clouds will not adopt Sky Computing, so it has no chance of succeeding.*
**Response:** An intercloud broker can use the services in any cloud, no matter what each cloud provider thinks about Sky Computing. Some clouds might tailor their business practice to Sky Computing more than others (e.g., by supporting more services in the compatibility set, and offering reciprocally-free data peering), but none of them can block the emergence of Sky Computing. In addition, as we observed in Inset D, there are already cases where an intercloud broker could bring significant benefits to customers. The only question is how large this market becomes, not whether clouds will allow it.

**Objection**: *Users will prefer to run their jobs in a single cloud because (i) there will always be a dominant cloud that will provide faster and/or cheaper services than others and (ii) it is harder (and therefore more expensive) to design workloads that can run in multiple clouds.*
**Response:** When we look at recent history (see Figure 4), the market share for the most dominant cloud (AWS) has been relatively stable (around one third of the market). The collective market share of the next two clouds (Azure and Google) has been growing (now slightly under one third of the market), with the share of the next ten clouds relatively stable at roughly 20% of the market, with the market share of the smaller clouds falling to roughly 15%. Thus, we see no evidence to expect a single cloud to capture the majority of the market. It is possible that in the future there will be several large clouds with the market share of the long tail of smaller clouds continuing to shrink in size, but this still represents a fairly competitive market and the intercloud brokers will only intensify the competition. In addition, the presence of intercloud brokers will help the smaller clouds compete, which might reverse the recent trend of their collective market share. Yet another factor that could intensify the competition is that users might proactively adopt a cross-cloud strategy, regardless of short-term cost implications, for a variety of reasons, including regulatory requirements, the need for low latency and/or higher reliability, and fear of lock-in.

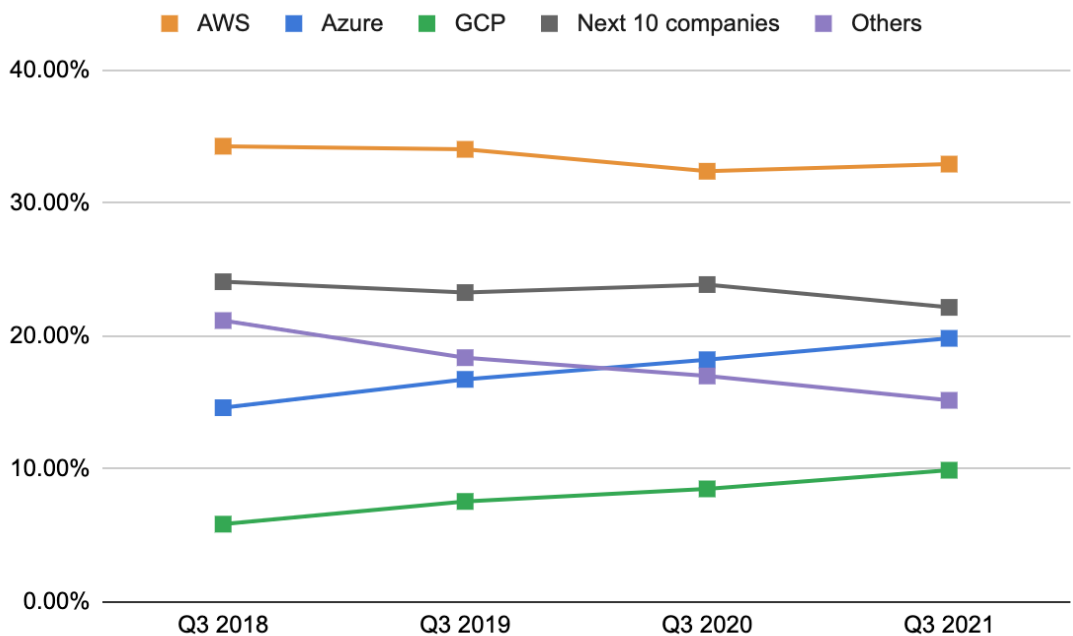

**Figure 4**: *Cloud market share as of October 2021 (see here).*

As to the second part of the objection, the individual developer does not have the task of porting jobs to multiple clouds, merely ensuring that they can use the Job API. While this is not trivial, it does not require mastering cloud-specific details, and there are many tasks that fit into a DAG model, as evidenced by the popularity of platforms such as Apache Airflow that is commonly used to build data and ML pipelines. For such tasks, an intercloud broker can automatically run them across multiple clouds if that best meets the user's needs. More fundamentally, we expect that the job API will rapidly be extended to additional use cases, and there too the Sky will handle the partitioning over multiple clouds.

**Objection**: *Isn't most job placement driven by data gravity, in which case everything you said is moot?*

**Response:** In the current cloud ecosystem, users tend to place all, or large segments, of their workloads on the same cloud, because running on multiple clouds remains a challenge. Thus, they naturally try to colocate this workload with their data. With Sky Computing, different jobs, and even different stages of the same job, can run on different clouds. There are many jobs that do not require much data (e.g., ML serving), and these could be moved to clouds for price or performance reasons. In addition, for many compute-intensive jobs like ML hyperparameter tuning, the delay of data transfer will typically be small compared to the length of the computation, so the remaining concern would be the cost of that transfer. With data transfer services like [Cloudflare's R2](Cloudflare's R2) mentioned earlier, and eventually the spread of reciprocally-free data peering, we think such costs will drop significantly. Finally, note that for an application whose computational complexity is superlinear in its input size, once the input's size exceeds a certain threshold, both the application's runtime and cost can be minimized by running the application on a cloud that provides faster or cheaper computation.

**Objection**: *Companies don't care about small improvements in price or performance (say, anything less than 20%), so what makes you think this can succeed?*
**Response**: First, as stated in the introduction, there are many reasons besides price or performance that are driving the need for workflow portability (e.g., data sovereignty), so even if the cost savings were insignificant there would still be reasons for Sky to emerge. Second, the objection's line of reasoning applies when it requires significant effort or expense to achieve these savings. However, in Sky Computing an intercloud broker automatically identifies the optimal placement, so the user does not need to spend time investigating tradeoffs. The only change on the part of the user is to convert jobs into Sky's job API. There are classes of jobs where this is extremely easy (e.g., the DAG example we described) and we expect that this will soon extend to other jobs as the job API evolves. However, even if the cost of developing workloads that fit Sky's job API is large, we expect that the savings will offset the costs for large companies. Indeed, a company that spends $100M/year on cloud would achieve $20M/year in savings given a 20% cost improvement; this is enough to pay a team of several tens of engineers, so we expect that such companies would devote significant resources to converting their workloads.

**Objection**: *Customers won't give up their volume deals, so they will keep their workloads where they presently are.*
**Response**: In the beginning, we see Sky Computing appealing most to smaller customers whose workload can be more easily decomposed into one-time requests and who do not presently have large discounts binding them to a particular cloud provider. Once Sky Computing gains traction, then brokers themselves may be able to negotiate volume discounts (just as travel sites do) so that all users, not just large ones, enjoy significant discounts.

## S4.2: Risks

**Risk**: *It is possible that the Sky never creates an effective market due to collusion (clouds paying brokers to guide jobs to them, or colluding among themselves), failure to create accurate* *enough catalogs (due to clouds not providing accurate service or pricing information), or predatory pricing (large clouds lowering prices to force smaller clouds out of business).*
**Response**: We hope that the combination of antitrust laws in various countries and global competitive pressures between brokers is sufficient to deal with most collusive behaviors, and to create reasonably accurate catalogs. Predatory pricing would be a major concern if the smaller clouds were startup companies with limited budgets. However, many of them have large companies (Google, Alibaba, IBM, etc.) behind them, so predatory pricing is unlikely to be effective.

**Risk**: *The most worrisome risk is that Sky never reaches escape velocity, and instead remains only a niche aspect of the cloud ecosystem.*
**Response**: We cannot dismiss this fear, as it represents a real possibility. However, we can try to avoid this fate by creating examples of popular job paradigms (i.e., job submission templates that could be usable for many use cases) that might catch the attention of larger users if we can demonstrate significant savings (in time or money). We will start with machine learning jobs from the research community (see [Inset E](#)), since their training and serving requirements are similar in construction to more commercial applications. Nonetheless, this risk remains the biggest question mark for Sky Computing.

**Risk:** *New decentralized computing models ("Web3") will disrupt the current cloud ecosystem in ways that are incompatible with the Sky vision.*
**Response:** We believe Web3 is largely complementary to Sky computing. On one hand, existing clouds can provide Web3 services as part of the compatibility set. Indeed, clouds like AWS and IBM are already providing blockchain services [[AWS-blockchain](#), [IBM-blockchain](#)]. On the other hand, if broadly adopted, Web3 can have a significant impact on several key components of Sky computing such as identity management and billing (e.g., creating a cryptocurrency-based market for buying cloud services). However, the jury is still out on whether Web3 could efficiently support data and compute intensive services such as Big Data analytics, distributed training, and model inference.

## S4.3 Additional Opportunities

The previous text listed many advantages of Sky Computing, but here we list some more surprising opportunities that Sky Computing could provide.

**Opportunity**: *Sky Computing could become used by the clouds themselves.*
**Explanation**: Large clouds have multiple availability zones (or regions).[15] In addition to providing guidance to users about how to place their jobs for resilience, these availability zones may have different offerings (e.g., TPU support is only available in some of Google's regions). Users will find the Sky useful for navigating these differences in offerings (they merely put TPU support as one of their requirements, rather than having to research what clouds, and which availability zones within those clouds, offer the required resources) and relieve the user of having to manually choose which zones to use for resilience (they merely specify that portions of the job have to run in different availability zones). Thus, even for large clouds that do not themselves orient themselves to Sky Computing, their users may find Sky Computing a convenient way of taming that cloud's internal complexity. If this becomes commonplace, even large cloud providers might see the job API as a way of reducing customer-specific configurations, thereby simplifying their own internal operations.

---

[15] AWS groups resources in a certain geographic region in "availability zones," while Azure and GPC use "regions."

**Opportunity**: *Sky Computing can direct jobs towards cleaner clouds.*
**Explanation**: The intercloud brokers can use a wide variety of criteria to guide job placement, and one that is becoming increasingly important is the carbon footprint of clouds. As recently noted in [Patterson et al.], the choice of DNN, datacenter, and processor can reduce the carbon footprint (as measured by $CO_2$ equivalent emissions*)* of large neural network training by up to three orders of magnitude. It would be easy for environmentally responsible companies to include this metric in their criteria; governments could also begin requiring certain cloud emission standards for work done on government contracts.

**Opportunity:** *Research can speed the adoption of Sky Computing, and leverage its existence.*
**Explanation:** There are several research areas that could speed the adoption of Sky Computing, mostly concerning how to overcome the inevitable heterogeneity of clouds and their services. For instance, service catalogs could be made far more accurate with tools that verify whether a cloud has a compatible implementation of a given service. These tools could be implemented by third parties and made available as open source, so everyone can inspect their accuracy and improve them.

At the core of the Intercloud broker lies an optimizer that—given the application's requirements and the available services—computes a physical execution plan for the application and then selects the appropriate compatible services across the clouds on which to run different application stages.[16] To make the best placement decisions, the optimizer also needs to accurately estimate the cost of running each application stage on a particular cloud service. Given the diversity of Sky applications, we will need to develop sophisticated cost models or efficiently estimate the cost at the runtime, e.g., run the same application stage on different clouds, monitor the progress of each run, and then terminate the slower or the more expensive runs.

Another critical aspect of Sky Computing is coming up with a framework to specify the applications' dependencies, constraints, and preferences. For batch applications like ML and data pipelines one possibility would be to use an Airflow-like specification. This specification could capture the execution and data dependencies between different stages of the application, as well as the set of services each stage could use. Furthermore, we could use a declarative language for specifying applications preferences (e.g., optimize for cost or latency) and constraints (e.g., process data within Germany's borders). We expect these specifications to evolve as the number and diversity of the applications supported by Sky Computing is going to increase.

---
[16] We use "stage" to define a component in the application's computation graph that cannot be split across services or clouds. Examples of stages include training, serving and extract-transform-load (ETL).

Running different stages of the same application on different clouds would require moving the data between clouds. If the amount of data is large, this can be expensive (due to the egress fees) and incur high latency. Reducing the transfer cost and latency can be critical to improving the performance and even the feasibility of Sky applications. Possible techniques to achieve these goals include overlay routing (e.g., instead transferring data from AWS West to GPC East directly, using either GCP West AWS East as waypoints), using multiple connections, leveraging compression to trade computation for the data being transferred, and using copy-on-write to avoid transfering unneeded data.

In addition, we will need new debugging, monitoring, and profiling tools that can identify bugs and performance bottlenecks in a Sky application and accurately attribute them to a particular cloud. One challenge here is to navigate the complex privacy regulations across different geographic regions and clouds.

Finally, it will be necessary to provide a clear understanding of data consistency for jobs that run across a variety of services. Since we expect that much of the data to remain stored in existing cloud storage systems (e.g., blob stores such as S3), we will need to provide bolt-on consistency [Bailis et al 2013] on top of these storage services to avoid data replication.

These are just a few examples of an active research agenda that will be needed to help Sky's adoption. Perhaps more interestingly, once established, the focus can move towards how to best leverage Sky Computing. For instance, how can one best build applications that span multiple datacenters? And can we use such approaches to improve application security and reliability by running them across multiple clouds by, for example, distributing trust? These approaches represent an exciting new area of research that can allow users to avoid being over-reliant on the security or reliability of a single cloud.

## S5: Conclusion

One might read this document thinking that a full transition to Sky Computing will be much more complicated than we have described, and thus our arguments are not grounded in reality. We completely agree with the former assertion, but strenuously dissent from the latter one. There are many complicated issues—such as designing more sophisticated solutions to various broker components (e.g., the job API, IMA and other security issues, and service catalog validation and maintenance)—that must be more adequately addressed before Sky Computing can handle most of the current cloud computing market. However, we contend that we *need* not and *should* not address these complications now.

Good system design focuses on problems that address current bottlenecks. The reason we think that the many issues we have not yet sufficiently addressed can and should be safely put off until later is because, as we have argued, there are enough "easy but usable" cases that we can start with. Creating an intercloud broker that can handle these cases will, we think, kick off a process that will take on a life of its own, creating more cloud compatibility and allowing the Sky to control a larger portion of the cloud market. During this early stage of Sky Computing we will learn much about both users and clouds, particularly how they might change their expectations and practices due to Sky's existence. For instance, over time users may start adapting their jobs to fit the current job API, and clouds might start altering their certain operational practices (such as pricing) to be better supported by brokers. Therefore, we should not despair that Sky Computing cannot handle everything clouds are used for today. The right time to address the more complicated issues will be when they first become a significant bottleneck in further Sky Computing adoption because then, and only then, will we know which complications are most important to address and which have become largely moot.

Despite the agile approach described above, we should be clear that what we are proposing is nothing short of a revolution in an important technology ecosystem. Unusually, it is a revolution based not on a major change in an underlying technology, but based only on the introduction of a two-sided market. There are good reasons to doubt that one can make a transformative change in a vital and vast technology ecosystem with such a seemingly small change. However, we don't expect this change to come rapidly, but rather through the gentle but relentless pressure of the ensuing market forces.

The resulting transformation, changing the basic usage model of the cloud from gathering and configuring resources in a specific cloud to merely specifying a job description, will expand cloud usage by: making the cloud easier to use, increasing the pace of technical innovation through the rise of specialized clouds, including edge computing and on premise computing within the cloud computing paradigm, and enhancing security and resilience via cross-cloud deployments.

At this point, however, these are just predictions, which many will doubt. We realize that every successful technology revolution must make the perilous passage from the seemingly impossible to the obviously inevitable. We view this white paper as only the first step in this journey; we don't ask that readers accept Sky Computing as inevitable, merely as not impossible. And we invite interested readers to help us on this quest.

Inset A: a description of how Sky is different from existing multi-cloud solutions.

Multi-cloud is emerging as a rapidly growing trend in cloud computing. However, so far there is little consensus on what "multi-cloud" means. Here, we differentiate between two current types of multi-cloud solutions, and explain how Sky is different from both of them.

|  | **Partitioned multi-cloud** | **Portable multi-cloud** | **Sky (transparent multi-cloud)** |
| --- | --- | --- | --- |
| Same app running on different clouds? | No | Yes | Yes |
| Cloud transparent? | No | No | Yes |
| Universal APIs (do all clouds provide the same APIs)? | No | Yes | No |
| Deep APIs (are APIs at different levels)? | Yes | No | Yes |

**Table 1:** *Comparison between existing types of multi-cloud and Sky.*

**Partitioned multi-cloud**: According to multiple recent reports [Pablo 2021, HashiCorp 2021, Accenture 2021], the majority of enterprises are already multi-cloud or at least have a multi-cloud strategy. However, these reports consider a company to be multi-cloud when different corporate teams run their workloads on different clouds or on-premise clusters. For example, one team could use Azure's Synapse to perform data analytics, another team could use AWS' SageMaker for running its ML workload, and yet another team could use GCP's Vertex AI for running another ML workload. We call this **partitioned multi-cloud** since, while a company uses multiple clouds, each workload runs on a single cloud.

**Portable multi-cloud:** A more advanced form of multi-cloud is when the same application runs on different clouds. Many third-party cloud applications (e.g., Confluent, Databricks, Snowflake, Trifacta) currently run on different clouds. As noted in section 1, this allows them to increase their addressable market and provides them with a competitive advantage over cloud-specific offerings. Companies like VMware provide tools and platforms to help companies with developing cloud portable applications [VMware]. However, portability should not be confused with transparency. These applications do not abstract away the clouds. Indeed, most of these applications request the user to first select a cloud and a region before creating an account.

This category also includes efforts to provide a uniform interface (APIs) across multiple clouds. Typically, such an interface is low level, such as cluster orchestrators (e.g., Kubernetes), virtual machines or container images, network virtualization, and so on. Recent examples are Google Anthos and Azure ARC. Another example is a previous "Sky computing" proposal that aimed at providing an uniform infrastructure-as-a-service for applications such as high-performance computing (HPC) applications [Katarzyna et al. 2009, Monteiro et al. 2011]. Similarly, VMware has popularized the idea of InterCloud, again as a uniform layer across clouds [Jin, 2010]. These still require the user to choose the cloud, but they make portability easier.

**Sky**. In contrast to existing multi-cloud solutions, Sky abstracts away the clouds. When running a Sky application, it is *transparent* to the user which clouds the application runs on. In addition, the set of interfaces are *heterogeneous* and *deep*. It is heterogeneous in that the interfaces belonging to the compatibility set do not need to run on all clouds. It is deep in that the compatibility set includes not only low level interfaces, but also high level interfaces such as analytics and ML.

Another relevant effort is Gaia-X, a large European Union project. Its goal is to develop a secure data platform based on open standards that "links" many cloud services together. This would enable transparent access to data across clouds, while still allowing data owners to control where their data is stored. Gaia-X is complementary to Sky as it focuses on transparent data access, while Sky focuses on providing transparent access to various public cloud services without requiring cloud cooperation.[17]

---

[17] In contrast, Gaia-X assumes the existence of a contract framework between cloud providers and customers [Policy Rule Document].

Inset B: Detailed Description of the Example.

Figure 5 shows a ML pipeline consisting of three stages:
- *Confidential data processing*: remove sensitive information from raw data using Intel SGX hardware enclaves. We use the Amazon Customer Reviews Dataset and treat it as if it contained personally identifiable information (PII) and thus must be processed securely. To remove sensitive data, we run Opaque on an SGX-enabled instance to filter on a column (i.e., the filtered-out information is assumed sensitive), and output only the review texts and star ratings. The size of the output dataset is 1 GB.
- *Training*: fine-tune BERT, a popular natural language understanding model, on the preprocessed (nonsensitive) data. This model predicts a rating given a review text. We fine-tune the model for 10 epochs.
- *Inference*: use the model to classify 1M new reviews.

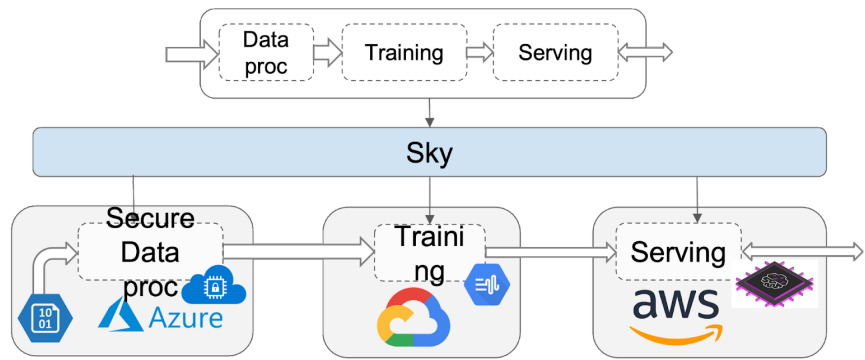

**Figure 5**: *An ML pipeline where different stages can run on different clouds.*

Our objective is to minimize the total cost while respecting the confidential computing requirements. We assume we have access to the three major public clouds: Azure, AWS, and GCP.

|  |  | proc | train | infer | egress | **Total** |
|---|---|---|---|---|---|---|
| **Time (hr.)** | **Azure** | 0.6 | 8.4 | 1.8 | – | 10.8 |
|  | **Sky** | 0.6 | 4.0 **-47%** | 1.1 **-38%** | 0.03 | 5.7 **-47%** |
| **Cost ($)** | **Azure** | 0.8 | 103 | 1.4 | – | 105 |
|  | **Sky** | 0.8 | 39 **-62%** | 0.3 **-78%** | 0.1 | 40 **-61%** |

*Table 1: Time and cost savings of running the pipeline in Figure 5 over Sky instead of a single cloud (i.e., Azure).*

Since the first stage requires instances based on SGX, we are restricted to use Azure Confidential Computing. We assume the data is initially stored in the Azure Blob Store. For training and serving we have more options. For instance, all clouds have instances with four Nvidia V100 GPUs which can be used for training, and instances employing Nvidia T4 GPUs which can be used for serving. Such instances are available on all three clouds.

We assume a cost-based optimizer that computes a physical plan for our pipeline. Despite the many cloud options providing Nvidia GPUs, this optimizer chooses TPU instances in GCP for training and the Inferentia service in AWS for serving. As a result, using Sky, the cost of running this pipeline is 61% lower than executing the same pipeline on a single cloud. (When we are using a single cloud, that cloud must be Azure since the first pipeline stage has to run on Azure.) The cost savings take into account the egress costs of moving the data from Azure to GCP for training, and the model from GCP to AWS for serving. This plan not only reduces the cost significantly, but it also almost halves the running time when compared with running the pipeline in Azure. Again, these improvements are after factoring in the overhead of transferring the data across clouds. Finally, while the cost savings in absolute terms are not large, note that the cost of training for more sophisticated models can be huge, and the cost of serving scales linearly with the number of queries. For example, it was reported [Shaabana, 2020] that the cost of training GPT-3 is equivalent to $12M at the current GPU/CPU prices. In such cases we expect Sky to provide cost savings in millions of dollars.

Inset C: OSS projects that are currently hosted across multiple clouds.

Over the past decade, more and more open-source projects have become de facto standards at different layers of the software stack. Furthermore, many of these projects can be relatively easily deployed in the cloud setting, and many clouds and third parties offer managed services of these projects. These trends make these open-source projects natural components of the compatibility set. Next, we give a (hopelessly incomplete) list of such projects.

- **Infrastructure**
  - **Linux and Docker:** These are the de facto OS and application packaging standard for single-node applications. Today, every cloud provides Linux-based images and managed docker registries.
  - **Kuberenetes, Terraform**: These projects are emerging as de facto standards for cluster orchestration and managing distributed applications. All major clouds provide Kubernetes as a service; e.g, Azure provides AKS, AWS provides EKS, and Google provides GKS. In addition, other third parties, such as RedHat and VMWare, provide hosted Kubernetes services (e.g., [OpenShift](), [Tanzu]()). Finally, HashiCorpprovides hosted services of its Terraform software on all major clouds.
- **Data processing:**
  - **Apache Spark**: This is the de facto standard for big data processing. Not only can one run Apache Spark on Kubernetes or natively on multiple clouds, but many clouds and third parties provide hosted versions of Apache Spark. For example, Azure provides Apache Spark both as part of HDInsight and Synapse. AWS and GCP provide Apache Spark as part of EMR, and DataProc, respectively. On top of that, Databricks provides managed Apache Spark on all major clouds.
  - **Apache Kafka**: This is the de facto standard for scalable data ingestion and stream processing. Again, different clouds provide managed Apache Kafka services, e.g., HDInsight on Azure, Managed Streaming for Apache Kafka (MSK) on AWS, and Bitnami on GCP. Finally, Confluent provides a hosted Apache Kafka service on all major clouds.
  - **PostgreSQL, MySQL**: These are very popular single-node database offerings. All major clouds provide hosted versions of these databases, e.g., [Free Database on AWS](), [Azure's PostgreSQL](), [GCP's Cloud SQL]().
  - **MongoDB, Cassandra**: These are popular NoSQL distributed databases. MongoDB and DataStax provide managed versions of these databases on multiple clouds.
- **Machine Learning**:
  - **TensorFlow, PyTorch, [Apache MXNet](), [MLFlow]()**: These are the dominant ML libraries. Major clouds provide instance images for most of these libraries, including instances using accelerators like NVIDIA's GPUs and TPUs. In addition, there are docker images available for these libraries.
  - **[PyTorch LIghtning](), [Horovod](), [Ray]()**: These are distributed frameworks targeting ML workloads. PyTorch Lightning and Horovod focus on distributed training, while Ray is a general-purpose distributed framework that supports several scalable

ML libraries, including RLlib, Tune, and Serve. One can instantiate these frameworks on every major cloud or use the managed versions provided by clouds' ML platforms such as SageMaker or Azure ML. Finally, companies like Anyscale and grid.ai provide hosted versions of Ray and Pytorch Lightning, respectively, on multiple clouds.
- **Storage**:
    - [Parquet](#): This is a columnar storage format for data lakes. It is supported by all major clouds' blob stores. This makes it possible for data processing and ML frameworks to read and write data using the same format on different clouds.
    - [Delta Lake](#), [Apache Iceberg](#), [Apache Hudi](#): These are tabular data formats for data lakes that provide atomic updates, consistent updates, and eliminate metadata bottlenecks. All these formats are available on all clouds.
    - [Delta sharing](#): This is an open-source protocol that allows sharing the data in Delta format. This enables applications sharing data across clouds. For example, an instance of Apache Spark running in AWS can use Delta sharing to read data stored in Azure.

While virtually all of these open-source projects run on multiple clouds, not all offerings are the same. For instance, sometimes clouds might fork an open-source project and host the forked version of the project [Root et al, 2019]. This could lead to unintended fragmentation. However, in many cases the users can pick the desired *version* of an open-source project on their cloud of choice. For example, users can run any of these open-source projects on any cloud, at the cost of managing the deployments themselves. Furthermore, many managed services also allow users to explicitly select the particular version of the open-source project they want (see Figure 6).

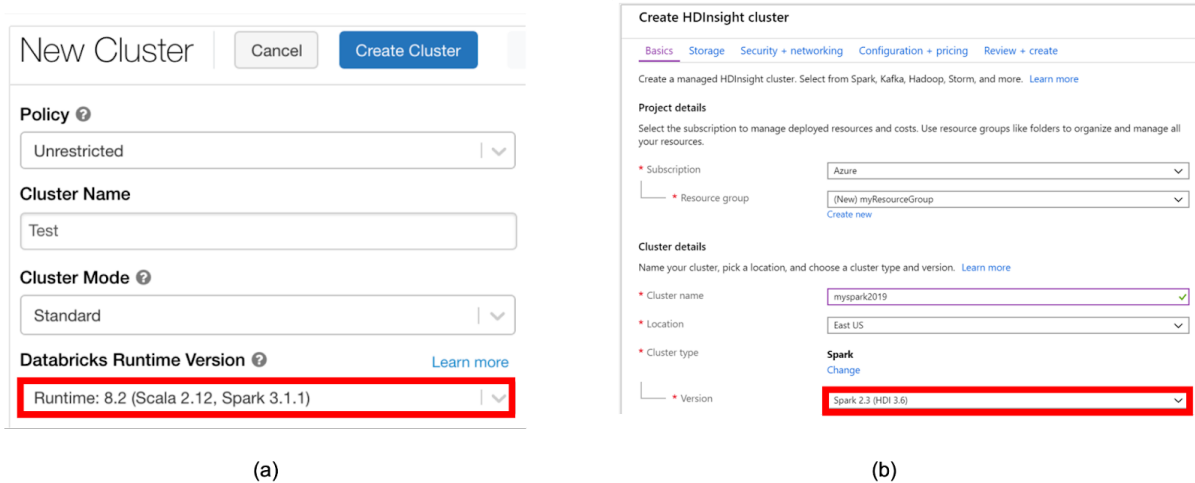

(a)            (b)

**Figure 6**: *Examples of cloud services allowing the user to select the desired Apache Spark version. (a) Databricks. (b) Azure HDInsight*.

Inset D: Examples of jobs that can run on multiple clouds with significant price/performance and other advantages for doing so.

There are many motivations to break a large computation across multiple clouds, but primarily they break down into five broad categories: cost savings, governance, performance, scale, and security goals. Inset B offers a Machine Learning example. Here we sketch several more scenarios that exercise these issues. Our goal here is illustrative, and as such we leave to future work the challenge of quantifying these benefits precisely at some moment in time, as we did in Inset B.

*Scenario 1 (cost savings): Mixed pricing for high-variance workload management.* Here we envision a retail chain allocating compute and storage (e.g. transactional databases, web servers, application servers, queuing services, prediction servers for recommendations, etc.) for online shopping. Based on trends from previous years, the chain has a good sense of its typical daily compute and storage needs, which—excepting seasonal effects—are fairly stable. The retailer uses a broker to contract an annual resource reservation with Cloud A for this workload. Meanwhile, the retailer can predict much of the seasonality, and uses the broker to negotiate advance reservations for daily or weekly "predicted burst" resources for those dates with Cloud B. Finally, the retailer is determined to have an ability to handle "unpredicted bursts" via short-term allocation of compute and storage on spot markets, possibly run by special brokers, using whatever cloud is currently offering a deal. Because all three of these rental regimes have different timescales, the best deal may come from three or more different clouds at any given time.

*Scenario 2 (Governance): Data Residency, Data Sovereignty, Geo-Proximity.* It is common in many nations to either incentivize or require data *residency:* that the data generated within the country (say customer data) remain stored within the country. A separate issue is data *sovereignty* [Guseyva 2021], which requires that the data be managed in accordance with national laws, e.g. regarding privacy. Any multinational corporation has to comply with these regulations. As one example, Canada has strict residency rules, but as of today AWS has only one region in Canada, and none on the west coast. Azure has regions in both central Canada and the west coast. So in order to offer good performance under data residency law, many AWS customers use Azure for low-latency west-coast services in Canada. While this may change over time (e.g. AWS plans a west-coast presence in Canada in a couple years), these changes are localized, slow and expensive; businesses will embrace workarounds in the interim and that will change the market dynamics.

*Scenario 3 (Performance): Multi-tier storage for performance.* Here we discuss a sketch for a natural Sky storage service, inspired by designs like Log-Structured Merge (LSM) trees [O'Neil et al. 1996], which might naturally be distributed across clouds for performance reasons. This is an example of a "built-for-the-sky" service abstraction, composed of sub-services running across clouds. The idea is to provide the performance and consistency guarantees of a low-latency database via the use of three services. (a) The first is an *Online Database* that keeps recently-ingested data in a main-memory database for rapid, low latency reads and writes. All requests for individual data items go first through this tier, and achieve whatever consistency guarantees this tier offers. (b) The second service is a *Near-line Database* that maintains the archive of less-recently-updated data. This database grows relatively slowly, over time, via cold data being bulk-loaded out of the second tier. It does not need fine-grained concurrency control since updates happen in batch, but it does need good index read latency; search indexes (e.g. Elastic) or SQL data warehouses may be appropriate for this tier. Only a fraction of queries (on cold data) pass through the cache to this layer. (c) The third service is a streaming *merge* service that ingests batches of data to be stored (either from the online database or from external sources), and transforms them for efficient bulk loading into the Near-line Database, taking advantage of that system's mechanisms for bulk updates (which may afford relatively strong consistency). Notably, *each of these services has different requirements for latency, bandwidth, memory and persistence*; as a result the best choice of cloud for each service may well be different.

*Scenario 4 (Scale): Resource aggregation.* The demands of training deep neural networks (DNNs) are increasing much faster than the capabilities of a single processor (e.g., GPU), so distributed training is becoming a necessity. Today's largest DNNs, such as GPT-3 [Brown et al. 2020] or Wu Dao 2 [Romero 2021] need hundreds or even thousands of GPUs to train. Furthermore, if we want to perform hyperparameter tuning on these networks, the resource requirements can easily balloon by orders of magnitude. Thus, it should come as no surprise that some companies are already finding it difficult to get enough GPUs on a single cloud to train their workloads. In fact, some companies and researchers are already aggregating resources across clouds using ad-hoc solutions. For example, a group of researchers from the San Diego Supercomputer Center used such a solution to allocate 50K GPUs across the top three clouds—AWS, Azure and GCP—to accurately simulate neutrino interactions [Sfiligoi 2019]. Sky can solve, or at least alleviate, this challenge by transparently running the user's workloads across multiple clouds. For example, Sky can run various trials of a hyperparameter tuning workload on different clouds, and then aggregate the results before returning to the users.

*Scenario 5 (Security): Multi-cloud security.* Assuming each cloud is a different trust domain, it is possible to develop secure multi-cloud systems and services to guarantee data confidentiality as long as the attacker fails to compromise at least one cloud. One such example is Dory [Dauterman et al. 2020], an encrypted search system that leverages multiple clouds as multiple trust domains. In a nutshell, these systems split the trust across multiple clouds so that a malicious attacker needs to compromise all clouds to obtain the plaintext data. This is analogous to building highly available distributed systems that are available as long as at least a server is still up. Sky Computing could considerably simplify building such applications and services. Furthermore, once built, these services can become part of the compatibility set thus enriching Sky computing functionality.

*Scenario 6 (Multiple Goals): Cloud augmentation of edge-devices.* Cloud computing can allow Internet of things (IoT) along with sensors, motors, and robots with limited on-board compute capabilities to increase performance by sharing data and using ML to learn and update perception and control policies; e.g., to recognize pedestrians and to dynamically compute motion trajectories and grasps [Kehoe et al. 2015, Nazari et al. 2020, Seredynski 2021]. Such policies can benefit from streams of new data from differing physical environments and hardware platforms. In many IoT and robotics applications, computing requirements vary between long periods of inactivity to intensive short bursts of computing to stochastically optimize motions and recover from failures. Also, in many of these cases, latency is critical. Sky Computing could accelerate IoT and robotics by providing more cloud options that can help with latency (e.g., choose a cloud with a nearby datacenter), security (e.g., choose a cloud employing hardware enclave instances) and cost (e.g., choose the cheapest cloud). The challenge here is that these constraints are very location-dependent (based on where the robot/sensor is), so manually configuring each individual case is infeasible, but a natural fit for intercloud brokers. For example, Sky Computing could facilitate automated and semi-automated driving by actively monitoring urban intersections, combining data from local cameras and sensors with data from approaching vehicles, nearby intersections, and historic data to time traffic lights and assist vehicles in awareness of pedestrians and congestion to increase safety, and reduce energy and delays. In this example, privacy-sensitive data (such as faces and license plates) could be pre-processed by Azure Confidential Computing before sending to GCP for training with TPUs, and then use the cloud which provides the best cost-latency tradeoff for inference. Furthermore, latency-sensitive processing could be handled by edge-computing clusters provided by regional vendors, such as 5G operators. Long-term archival requirements (for learning or legal reasons) could be handled by AWS glacier storage. As such, Sky computing could optimize performance and cost given latency constraints and security/regulatory constraints.

Inset E: Prototype.

The Sky computing proposal described in this paper is a daunting endeavor. To succeed, we believe two things need to happen: (1) the emergence of early proof points that illustrate the potential of Sky computing, and (2) a rapidly growing community. These two things are closely related: successful proof points will lead to the growth of the Sky community, and a larger Sky community will lead to more successful proof points.

We now briefly describe our plans to develop a first proof point, which is designed to validate our first conjecture that the existence of "*some easy but useful cases will be enough for Sky Computing to provide significant benefits to users.*" We do this by supporting training and hyperparameter tuning workloads for Berkeley's AI students starting with the RISELab students first. There are several reasons we have decided to start with this use case:
- *We understand the users*: The users are the students developing the prototype or their colleagues. Thus, the developers understand the pain points of users, which is critical in building any product or service as Christensen and his colleagues argued in their seminal "jobs-to-be-done" paper [Christinsen et al., 2016].
- *Significant problem*: AI workloads often dominate the cloud expenses of the labs due to their high computation requirements[18] and extensive use of hardware accelerators, such as GPUs and TPUs. At RISELab, we are spending over 75% of our cloud credits on running AI workloads.
- *Natural multi-cloud setting*: RISELab as well as other labs that conduct AI research already have academic credits from multiple cloud providers, including AWS, Azure, and GCP. However, since using each cloud requires customization of their scripts, many students end up using a single cloud. Our project aims to make it easier to utilize all available cloud credits.
- *Low data transfer costs*: One of the challenges in a multi-cloud setting is the egress cost of moving data across clouds. Fortunately, this transfer cost does not dominate the total cost of running these AI workloads because (i) the computation complexity of many AI workloads such as training are super-linear in the input size and (ii) many of the academic AI workloads use standard benchmarks (e.g., ImageNet) which can be easily replicated across clouds and then reused across many training episodes.

In addition, our application does not need to solve several general challenges inherent to Sky computing. This simplifies the problem we have to solve, and thus increases our chances of success.
- *No data security*: Since the vast majority of academic projects use public data, there is no need for sophisticated techniques to maintain data confidentiality.
- *No billing and accounting*: Our intended users already have accounts with the clouds providing academic credits. This means that our prototype needs to implement neither billing nor account management.

---

[18] https://openai.com/blog/ai-and-compute/

We are developing our prototype on top of the [Ray](#) open-source project. Ray is a general-purpose distributed framework that has a strong ecosystem of distributed libraries, including [Ray Tune](#), the most popular scalable hyperparameter tuning library. Ray and Ray Tune integrate with all major ML libraries, including TensorFlow and PyTorch. One of the primary reasons we use Ray is to leverage its powerful [Autoscaler](#) component, which automatically deploys and autoscales distributed Ray applications natively on AWS, Azure, and GCP, as well as on Kubernetes.

While our prototype doesn't need to support data security, accounting, and billing, there are still plenty of challenges it needs to address. In particular, our prototype implements the following functionality:
- *Simple service catalog*: We are developing a bare-bones catalog including only services required by our application. These include accessing data from the blob store of each of the major clouds, starting and stopping instances, and using the Ray Autoscaler.
- *Job API:* We are using a simple API capable of describing simple pipelines, similar to Apache Airflow. This API allows the user to specify the cloud on which the input is stored, as well as the type of accelerators supported by the application, e.g., GPUs and/or TPUs. For different accelerator types, the user is expected to provide the corresponding docker images. In addition the API also enables the user to specify its preferences, e.g., minimize the cost, as well as clouds can be used by the application.
- *Optimizer*: We are starting with a simple optimizer that aims to minimize the cost of running the application, including the egress cost, if any. We aim to optimize the cost not only across clouds (e.g., using TPUs on GPP vs GPUs on AWS) but also across different regions in the same cloud.
- *Executor*: The executor leverages Ray Autoscaler to launch and manage each stage of the pipeline as a single Ray application. First, the executor picks a cloud, and then tries to find a region with enough resources (spot instances, if possible) to run the stage. If it does not find any region with enough resources, the optimizer will try the next best cloud in terms of cost. The executor also implements a job queue which allows the user to submit multiple jobs, e.g., training episodes.

We have already implemented a first version of a prototype that we are testing internally. Our goal is to incrementally roll it out first to RISELab, then to other Berkeley labs such as BAIR, and then hopefully to other universities and organizations. Despite the relative simplicity of this application, we hope to learn valuable lessons about what users really value. In fact, we have already learned a few such lessons: even in the case of a single cloud, users value the ability to automatically find a region with enough resources to run their application; users want a common API to access the data in different clouds; users want a tool to copy the data efficiently across clouds; and users want the ability to detect when something is going wrong with their Sky application and pinpoint the problem. We are currently working on addressing all these problems in our prototype.

**Acknowledgements**. We would like to thank our many colleagues who have generously read the early versions of this paper and provided insightful feedback: Dirk Bergemann, Adrian Cockcroft, David Culler, Ian Foster, Mark Russinovich, David Tennenhouse, Marvin Theimer, Deepak Vij, and Matei Zaharia.